\documentclass[aps,prb,twocolumn,showpacs,superscriptaddress,amsmath,amssymb,floatfix,10pt]{revtex4-1}

\usepackage{graphicx}
\usepackage{amssymb}
\usepackage[toc,page]{appendix}
\usepackage{color}
\usepackage{comment}
\usepackage{hyperref}
\usepackage{float}
\usepackage{longtable}
\usepackage{booktabs}
\usepackage{threeparttable}

\usepackage{subfigure,amsmath,verbatim,moreverb}
\usepackage{tabularx}
\usepackage{adjustbox}
\usepackage{lipsum}
\usepackage{latexsym}
\usepackage{dcolumn}
\usepackage{amsmath}
\usepackage{dcolumn}

\newcolumntype{d}[1]{D{.}{.}{#1}}

\newcommand{\R}{\mathbf{r}}

\def\myq{q}
\newcommand{\be}{\begin{equation}}
\newcommand{\ee}{\end{equation}}
\newcommand{\bea}{\begin{eqnarray}}
\newcommand{\eea}{\end{eqnarray}}
\newcommand{\bean}{\begin{eqnarray*}}
\newcommand{\eean}{\end{eqnarray*}}

\begin{document}

\title{Adiabatic connection interaction strength interpolation method made accurate
for the uniform electron gas}
\author{Lucian A. Constantin}
\affiliation{Institute for Microelectronics and Microsystems (CNR-IMM),
Via Monteroni, Campus Unisalento, 73100 Lecce, Italy}
\author{Subrata Jana}
\affiliation{Department of Molecular Chemistry and Materials Science, Weizmann Institute of Science, Rehovoth 76100, Israel}
\author{Szymon \'Smiga}
\affiliation{Institute of Physics, Faculty of Physics, Astronomy and Informatics, Nicolaus Copernicus University in Toru\'n,
ul. Grudzi\c adzka 5, 87-100 Toru\'n, Poland}
\author{Fabio Della Sala}
\affiliation{Center for Biomolecular Nanotechnologies, Istituto Italiano di
Tecnologia, Via Barsanti 14, 73010 Arnesano (LE), Italy}
\affiliation{Institute for Microelectronics and Microsystems (CNR-IMM), Via
Monteroni, Campus Unisalento, 73100 Lecce, Italy}

\date{\today}

\begin{abstract}
The adiabatic connection interaction strength interpolation (ISI)-like method provides a high-level expression for the correlation energy, being in principle exact in the weak-interaction limit, where it
recovers the second-order Görling-Levy perturbation term, but also in the strong-interaction limit that is described by the strictly correlated electron approach. In this work, we construct the genISI functional made accurate for the uniform electron gas, a solid-state physics paradigm that is a very difficult test for ISI-like correlation functionals. We assess the genISI functional for various jellium spheres with the number of electrons Z $\leq$ 912 and for the non-relativistic noble atoms with Z $\leq$ 290. For the jellium clusters, the genISI is remarkably accurate, while for the noble atoms, it shows a good performance, similar to other ISI-like methods. Then, the genISI functional can open the path using the ISI-like method in solid-state calculations.
\end{abstract}

\maketitle

\section{Introduction}
\label{sec1}
The Density Functional Theory (DFT) \cite{hohenberg1964inhomogeneous,kohn1965self}, is exact in
principle, but in practice, the
exchange-correlation (XC) energy $E_{xc}[n(\R)]$ as a functional of the electronic density $n(\R)$
must be approximated. The XC functional should contain all the many-body quantum effects raised
by the electron-electron interactions beyond
the Hartree method. Nowadays, there are known important exact properties of
$E_{xc}[n]$ that have been used in the construction of many XC functional approximations, that are
classified on the so-called Jacob's ladder \cite{perdew2001jacob,perdew2005prescription}.
The first rung of the ladder is the Local Density Approximation (LDA)
\cite{kohn1965self,dirac1930note,perdew1992accurate} which has been constructed from the uniform
electron gas (UEG) model system\cite{loos16}. The UEG is one of the most important model systems for the XC
functional development, being a solid-state paradigm. In fact, LDA is still often used in solid-state
calculations because of its remarkable accuracy for various properties,
such as surface energy of transition metals \cite{patra2020improved} and work function
\cite{patra2017properties,singh2009surface}.

One of the most rigorous paths for constructing new XC functionals is via the
adiabatic connection (AC) method \cite{langreth1975exchange,gunnarsson1976exchange,savin2003adiabatic,
cohen2007assessment,ernzerhof1996construction,burke1997adiabatic,colonna1999correlation}, that
gives the exact XC energy as the following coupling-constant integral
\begin{equation}
E_{xc}[n]=\int_0^1 d\alpha\;W_{xc,\alpha}[n],
\label{eq1}
\end{equation}
where
\begin{equation}
W_{xc,\alpha}[n]=\langle\Psi_n^{min,\alpha}|\hat{V}_{ee}|\Psi_n^{min,\alpha}]\rangle-U[n].
\label{eq2}
\end{equation}
Here $U[n]=(1/2)\int d\R\int d\R'\;n(\R)n(\R')/|\R-\R'|$ is the Hartree energy, $\hat{V}_{ee}$ is the
Coulomb repulsion operator, and
$\Psi_n^{min,\alpha}$ is the antisymmetric wave function that yields the density $n(\R)$ and
minimizes the expectation value $\langle\hat{T}+\alpha \hat{V}_{ee}\rangle$,
with $\hat{T}$ being the kinetic energy operator, and $\alpha\ge 0$ the coupling constant (known also
as interaction strength). We note that $\Psi_n^{min,\alpha=1}=\Psi_n^{min}$ is the
interacting, exact ground-state wavefunction for density $n(\R)$ and $\Psi_n^{min,\alpha=0}=\Phi_n^{min}$
is the non-interacting KS wavefunction for density $n(\R)$ given in the form of Slater determinant.
Considering that the exchange energy functional is, by definition,
$$E_x[n(\R)]=\langle\Phi_n^{min}(\R)|\hat{V}_{ee}|\Phi_n^{min}(\R)\rangle-U[n(\R)],$$
then the adiabatic connection integrand $W_{xc,\alpha}[n]$ can be also written as
\begin{eqnarray}
&& W_{xc,\alpha}[n]=E_x[n]+W_{c,\alpha}[n],\nonumber\\
&& W_{c,\alpha}[n]=\langle \Psi_n^{min,\alpha}|\hat{V}_{ee}|\Psi_n^{min,\alpha}\rangle-
\langle\Phi_n^{min}|\hat{V}_{ee}|\Phi_n^{min}\rangle.\nonumber\\
\label{eq3}
\end{eqnarray}

The AC was often used in the construction of accurate hybrid
functionals \cite{burke1997adiabatic,kohn1996density,becke1993density,adamo1998exchange,UPBH}, and
especially
the most sophisticated, fifth-rung functionals, including the
G\"{o}rling-Levy (GL) perturbation correlation terms
\cite{gorling1994exact,gorling1993correlation,gorling1995hardness}, the random phase
approximation (RPA), and RPA-like  methods based on XC kernel approximations
\cite{dobson2002correlation,terentjev2018gradient,constantin2016simple,
corradini1998analytical,toulouse2005simple,richardson1994dynamical,
bates2016nonlocal,bates2017convergence,ruzsinszky2016kernel,dobson2000energy,
gorling1998exact,kim2002excitonic,erhard2016power,patrick2015adiabatic}.

In this work, we will focus on the high-level interaction strength interpolation (ISI) methods
\cite{seidl2000simulation,seidl2000density,perdew2001exploring,seidl1999strictly,gori2009electronic,
liu2009adiabatic}, that accurately
interpolate between the weak- ($\alpha\rightarrow 0$) and strong- ($\alpha\rightarrow \infty$)
interaction limits. Such methods have been intensively studied and tested
\cite{liu2009adiabatic2,
magyar2003accurate,sun2009extension,seidl2010adiabatic,mirtschink2012energy,
gori2010density,vuckovic2016exchange,fabiano2016interaction,giarrusso2018assessment,
vuckovic2018restoring,kooi2018local,zhou2015construction,
seidl2018communication,vuckovic2023density,fabiano2019investigation,smiga2022selfconsistent,daas2021noncovalent,
constantin2019correlation,smiga2020modified,vuckovic2017interpolated,vuckovic2017simple,daas2022large}.
Efficient implementation of ISI methods
are also available in public quantum-chemistry codes\cite{turbo2023}.

In the weak-interaction limit, the GL perturbation theory becomes exact, and
$W_{xc,\alpha}[n]$ is \cite{seidl2000simulation,gorling1993correlation,jana2020generalizing}
\begin{equation}
W_{xc,\alpha\rightarrow 0}[n]=W_0[n]+W'_0[n]\alpha+...+W^{(m)}_0[n]\alpha^m+....,
\label{eq4}
\end{equation}
where $W_0[n]=E_x[n]$ is the exact DFT exchange functional,
$W'_0[n]=2E_c^{GL2}[n]$, and $W^{(m)}_0[n]=(m+1)E_c^{GL_{m+1}}[n]$. In this work, as in most of the
ISI-like methods, we consider only the first two terms of the perturbation expansion.

On the other hand, in the strong-interaction limit, $W_{xc,\alpha}[n]$ behaves as
\cite{liu2009adiabatic}
\begin{equation}
W_{xc,\alpha\rightarrow\infty}[n]=W_\infty[n]+W'_\infty[n]\alpha^{-1/2}+W^{(2)}_\infty[n]\alpha^{-3/2}+...,
\label{eq5}
\end{equation}
where $W_\infty[n]$, $W'_\infty[n]$ can be in principle exactly calculated using the strictly
correlated electron (SCE) approach \cite{gori2009density,malet2012strong,friesecke2022strong}.
In practice, generalized gradient approximation (GGA) models
have been developed for $W_\infty[n]$ and
$W'_\infty[n]$\cite{seidl2000density,constantin2019correlation,smiga2022selfconsistent}.

Eq. (\ref{eq5}) shows that the $\alpha^{-1}$-term in the expansion should be zero. However,
the original ISI method \cite{seidl2000simulation} has a spurious $\alpha^{-1}$-term, that
was removed in the revISI \cite{gori2009electronic} and LB \cite{liu2009adiabatic} methods.

The total XC energy of ISI methods can be expressed as
\begin{equation}
E_{xc}[n]={\cal F}(E_x,E^{GL2}_c,W_\infty,W'_\infty) \label{eq:eexc}
\end{equation}
where ${\cal F}$ is a non-linear function of the ingredients
$E_x$ and $E^{GL2}_c$ and of the density functionals  $W_\infty$ and
$W'_\infty$.
Eq. \ref{eq:eexc} thus resembles (also from a computational cost point-of-view) the one of double-hybrid (DH) functionals\cite{goerigk2014double}: however, the former employs a non-linear dependence from the GL2 (or MP2) term  and do not diverges for systems with vanishing gaps\cite{fabiano2016interaction,smiga2020modified,turbo2023}, which is a clear superiority with respect to DH approaches.

However, the ISI methods have difficulties to recover the logarithmic singularity
of the UEG correlation energy per particle in the high-density limit ($\epsilon_c\propto \ln(r_s)$
when $r_s\rightarrow 0$, where $r_s$ is the bulk parameter), but all of them are accurate in the UEG
low-density limit (at $r_s \rightarrow\infty$) \cite{seidl2000density,harding2022approximate}. We recall that the high-density limit of the UEG correlation energy is exactly described by the RPA method \cite{perdew1992accurate}.

In spite of this limitation,
we construct an ISI-like method, named genISI, that is very accurate for the UEG correlation energy
when $r_s\ge 1$. Note that most materials are characterized by $1\le r_s \le 10$. Thus, one could expect increased accuracy of the proposed method.

The paper is organized as follows.  In section \ref{sec2}, we
present the construction of genISI XC functional. Computational details are described in Section \ref{seccd}, while in section \ref{sec3}, we report the
correlation energies results of genISI and other ISI-like methods for the UEG model system,
various jellium clusters, non-relativistic noble atoms and small systems for which the exact quantities ($W'_0$, $W_\infty$, $W'_\infty$) are known exactly. Finally, in section \ref{sec4}, we summarize our conclusions.

\section{the genISI XC functional}
\label{sec2}
We start the construction of genISI XC functional considering the UEG limit (denoted by UEG-ISI), where $W'_0\rightarrow -\infty$ and $W'_0/W_0 \rightarrow\infty$. In fact, in the UEG, the energy-gap vanishes and thus the GL2 correlation diverges.

\subsection{The UEG-ISI XC functional}
\label{sec21}
Let us consider the revISI expression \cite{gori2009electronic} for the UEG-ISI XC functional
%
\begin{equation}
W^{UEG-ISI}_{xc,\alpha}[n]=W_\infty[n]+\frac{b(2+c\alpha+2d \sqrt{1+c\alpha})}
{2\sqrt{1+c\alpha}(d+\sqrt{1+c\alpha})^2},
\label{eq6}
\end{equation}
so that the UEG-ISI XC energy is
\begin{eqnarray}
E^{UEG-ISI}_{xc}[n]&=&\int^1_0 d\alpha\;
W^{UEG-ISI}_{xc,\alpha}[n] \nonumber\\
&=&W_\infty[n]+\frac{b[n]}{d+\sqrt{1+c[n]}}
\label{eq11}
\end{eqnarray}
In Ref. \onlinecite{gori2009electronic} the parameters b, c, d were functions of $W'_0$, which tends to $-\infty$ for the UEG.
Here, instead, we define:
\begin{eqnarray}
&& b[n]=(W_0-W_\infty)(1+d), \nonumber\\
&& c[n]=b[n]^2 / [ 4 W'^2_\infty ].
\label{eq7}
\end{eqnarray}
The functionals $b[n]$ and $c[n]$ have been found such that
$W^{UEG-ISI}_{xc,\alpha}[n]$ has the
following properties:
\begin{eqnarray}
&& W^{UEG-ISI}_{xc,\alpha\rightarrow 0}[n]\longrightarrow W_0-
\frac{(W_0-W_\infty)^3(1+d)}{4W'^2_\infty}\alpha+\mathcal{O}(\alpha^2), \nonumber\\
&& W^{UEG-ISI}_{xc,\alpha\rightarrow \infty}[n]\longrightarrow
W_\infty+W'_\infty\alpha^{-1/2}+\mathcal{O}(\alpha^{-3/2}).
\label{eq8}
\end{eqnarray}
Then, $W^{UEG-ISI}_{xc,\alpha}[n]$ recovers the strong interaction limit of Eq. (\ref{eq5}), and
the first leading term of the weak interaction limit expansion of Eq. (\ref{eq4}). Of course, the
first perturbation term $W'_0[n]\alpha$ can not be fulfilled by $W^{UEG-ISI}_{xc,\alpha}[n]$,
because in the UEG limit $W'_0=-\infty$, such that $W^{UEG-ISI}_{xc,\alpha}[n]$ should be a
function only of $W_0$, $W_\infty$, and $W'_\infty$.

We also mention that, under the uniform scaling of the density $n_\lambda(\R)=\lambda^3n(\lambda\R)$,
with $\lambda\ge 0$ ($\lambda=\alpha^{-1}$), the functionals have the following scaling behavior
\cite{seidl2000density,gorling1993correlation}:
\begin{eqnarray}
&& W_\infty[n_\lambda]=\lambda W_\infty[n],\;\;\;W_0[n_\lambda]=\lambda
W_0[n],\nonumber \\
&& W'_\infty[n_\lambda]=\lambda^{3/2} W'_\infty[n],\;\;\;W'_0[n_\lambda]=
W'_0[n],
\label{eq9}
\end{eqnarray}
%
such that  
\begin{equation}
W^{UEG-ISI}_{xc,\alpha}[n_{\frac{1}{\alpha}}]=\alpha W^{UEG-ISI}_{xc,\alpha}[n].
\label{eq10}
\end{equation}
%


Note that all ISI-like functionals in the limit of $W'_0\rightarrow-\infty$ , can be written as
\begin{eqnarray}
E_{xc}[n]&=&W_\infty + W'_\infty F(\myq), \\
W_{\alpha}[n] &=& W_\infty +  (E_x-W_\infty) f( x ),
\end{eqnarray}
with
\begin{eqnarray}
    \myq&=&\frac{E_x-W_\infty}{W_\infty'} >0, \label{eq:qdef}\\
    x&=&\myq\sqrt{\alpha},
\end{eqnarray}
and $F(\myq)=\int_0^1 \myq f(\sqrt{\alpha}\myq) d\alpha$.
The functions $F(\myq)$ and $f(x)$ for common ISI-like functionals are reported in the Appendix A.
For the UEG-ISI we have
\begin{eqnarray}
F^{UEG-ISI}(\myq)&=&\frac{2 \myq (d+1)}{H(\myq)+2 d}, \\
f(x)&=&(d+1)\frac{ H(x)^2 +4 d H(x) +4} {H  (2 d+H(x)) ^2}, \\
 H(x)&=& \sqrt {4+(d+1)^2 x^2 }.
\end{eqnarray}
For small $q$ we have
\begin{equation}
F^{UEG-ISI}(\myq) \rightarrow \myq-\frac{\myq^3}{8} (d+1) +\mathcal{O}(\myq^5),
\label{eq:limueg}
\end{equation}
and thus
\begin{equation}
E_{xc}^{UEG-ISI}(\myq) \rightarrow E_x+ W'_\infty\left (-\frac{\myq^3}{8} (d+1) +\mathcal{O}(\myq^5) \right).
\label{eq:limuegxc}
\end{equation}

For one electron systems, if $W_\infty$ is computed exactly (which is the case for $W_\infty^{TPSS}$ and for the point-charge-plus-continuum (PC) model\cite{seidl2000density} for the hydrogen atom), we have that $q=0$ and thus UEG-ISI is correctly one-electron self-correlation free.
Actually for one electron systems, it should also be that $W'_\infty=0$, which can only be the case for a meta-GGA functional like TPSS: in this case  $q$ is undefined, but still $W'_\infty q^n $ vanishes for any n, and thus the UEG-ISI correlation is zero.
For other GGA models of $W_\infty$ and $W'_\infty$ see e.g. Refs.\onlinecite{smiga2022selfconsistent,constantin2019correlation} the genISI correlation of one electron systems is small but not exactly zero.

The only parameter of the UEG-ISI functional is $d$
and it is fixed as described in section \ref{sec31}
\subsection{The genISI XC functional}
\label{sec22}
To restore the $\alpha$-term of the weak interaction limit expansion of Eq. (\ref{eq4}) and
to preserve the UEG limit constructed above,
we consider the following expression for the genISI functional
\begin{eqnarray}
W^{genISI}_{xc,\alpha}[n]&=&W^{UEG-ISI}_{xc,\alpha}[n]+\nonumber\\
&& \frac{a[n]~p[n]~\alpha}{(1+r[n]~p[n]~\alpha)^3}
\label{eq12}
\end{eqnarray}
where
\begin{eqnarray}
a[n]&=&W_0\Big[ 1+\myq^3\frac{ W'_\infty }{4W'_0}(1+d) \Big],\nonumber\\
p[n]&=&W'_0/W_0, \nonumber\\
r[n]&=&m \Big( \frac{W_0}{W_\infty}\Big)^3,
\label{eq13}
\end{eqnarray}
and $m=18.0$ was fixed by fitting to the
correlation energy of the Hooke's atom with force constant $k=1/4$
\cite{taut1994two,constantin2019correlation}.
Considering the density scaling relations, we observe that
\begin{equation}
W^{genISI}_{xc,\alpha}[n_\alpha(\R)]=\alpha W^{genISI}_{xc,\alpha}[n(\R)].
\label{eq14}
\end{equation}
The genISI XC functional satisfies all the exact
conditions also recovered by revISI \cite{gori2009electronic} and LB \cite{liu2009adiabatic}
functionals, and additionally has an improved UEG limit.

The genISI XC energy is
\begin{eqnarray}
&& E^{genISI}_{xc}[n] = \int^1_0 d\alpha\;
W^{genISI}_{xc,\alpha}[n] \nonumber\\
&& = E^{UEG-ISI}_{xc}[n] +E^{add}_{xc}[n] \\
&& {\rm with \; }  E^{add}_{xc}[n]= \frac{a[n]p[n]}{2(r[n]p[n]+1)^2}.
\label{eq15}
\end{eqnarray}
 For small $\myq$ and small $E^{GL2}_c$ we have
\begin{eqnarray}
&& E_{xc}^{add} \rightarrow W'_\infty (\frac{\myq^3}{8}(d+1)  + \mathcal{O}(\myq^5) )+ \nonumber \\
&+& \left (1- \frac{m(d+1) W'_{\infty} ( \myq^3 +\mathcal{O} (\myq^4))  }{2 W_{\infty}}\right) E^{GL2}_c   + \mathcal{O}((E^{GL2}_c)^2) ,\nonumber \\
&&
\end{eqnarray}
which cancels the second term in Eq. \eqref{eq:limuegxc},
so that $E_x$ is recovered for a larger range of $q$.
Finally the total genISI XC energy behaves as
\begin{eqnarray}
&& E_{xc}^{genISI} \rightarrow
E_x+ W'_\infty \mathcal{O}(\myq^5) + \nonumber \\
&+&\left (1- \frac{m(d+1) W'_{\infty} (\myq^3 +\mathcal{O}(\myq^4)) }{2 W_{\infty}}\right) E^{GL2}_c  + \mathcal{O}((E^{GL2}_c)^2). \nonumber \\
\end{eqnarray}

For one electron systems $E^{GL2}_c=0$ and thus genISI is one-electron self-correlation free, as discussed in section \ref{sec22}.

Finally, we recall that the genISI total energy is not size-consistent (as all other interpolation formula), but a simple size-consistent correction can be added to it
\cite{vuckovic2018restoring}.

\section{Computational Details}
\label{seccd}
All the calculations for jellium clusters and atoms have been performed with a modified version of
the Engel code \cite{engel1993accurate,engel2003orbital}, using a radial numerical grid and  Kohn-Sham LDA orbitals and
densities. We consider 24 neutral jellium spheres with number of electrons $Z=$ 8, 18, 20, 34, 40,
58, 92, 132, 138, 186, 196, 254, 338, 398, 438, 440, 508, 556, 612, 638, 676, 758, 832 and $912$.
We also consider all noble atoms from $Z=$2 to 290 $e^-$.

Because the SCE and GL2 results are not available for such systems, we used some semilocal approximations of these quantities:
\begin{itemize}
\item $W_\infty^{TPSS}$ of Eq. (37) of Ref. \onlinecite{perdew2004meta} that is one the most
accurate models for $W_\infty$ (see Tables I-III of Ref. \onlinecite{perdew2004meta} and Table I of Ref.~\onlinecite{jana_semilocal_2023});
\item $W'^{MGGA}_\infty$ of Eq. (D16) of Ref. \onlinecite{seidl2000density} that is one the most
accurate models for $W'_\infty$ (see table II of Ref. \onlinecite{seidl2000density} and tables I and II of
Ref. \onlinecite{jana2022meta}).  We recall that the hPC GGA model for $W'_\infty$ is also very
accurate \cite{smiga2022selfconsistent};
\item $W'^{TPSS}_0$ of Eq. (A1) of Ref. \onlinecite{perdew2008density} that is one the most
accurate models for GL2 correlation energy (see Table S12 of Ref. \onlinecite{smiga2020modified});
\item $W_0=E_x$ the exact exchange computed with LDA orbitals.
\end{itemize}
We note that for jellium clusters and large atoms, the meta-GGA approximations of
$W_\infty$, $W'_\infty$, and $W'_0$ are expected to be accurate, such that their use in
ISI-like functionals should give realistic results.
\section{Results}
\label{sec3}

\subsection{UEG correlation energy}
\label{sec31}

The XC energy $E_{xc}=\int d\R\;n(\R)\epsilon_{xc}(\R)$ is just $E_{xc}=N\epsilon_{xc}(r_s)$ in case
of the spin-unpolarized UEG, where $N$ is the number of electrons and $\epsilon_{xc}(r_s)$ is the XC
energy per particle. We also note that $W_0=N (-3/(4\pi))k_F$, $W_\infty =An^{1/3}N$,
$W'_\infty=Bn^{1/2}N$, where $k_F=(3\pi^2n)^{1/3}$, $n=3/[4\pi r^3_s]$, $A=-1.451$ and $B=1.535$.
Thus $\myq=0.4641 n^{-1/6}$.
Substituting these expressions in the UEG limit of ISI-like functionals, we obtain their correlation
energies per particle $\epsilon_c(r_s)$. In the Appendix A, we show the expressions of ISI, revISI,
SPL \cite{seidl1999strictly}, and LB\cite{liu09} XC functionals and their UEG limits.

\begin{figure}[hbt]
\includegraphics[width=\columnwidth]{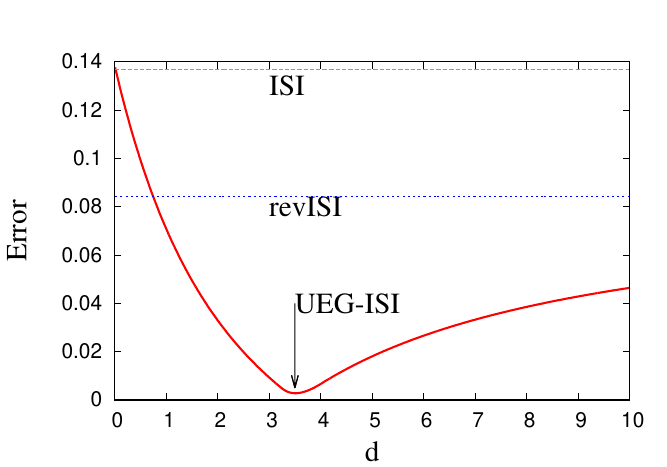}
\caption{
Error$=\int^{10}_1\;d\;r_s\;
|\epsilon^{approx}_c(d,r_s)-\epsilon^{exact}_c(r_s)|$
versus the parameter $d$. Also shown are the errors from ISI and revISI functionals.
}
\label{f1}
\end{figure}

The parameter $d$  of the UEG-ISI functional
was found by minimizing the expression
\begin{equation}
Error(d)=\int^{10}_1\;d\;r_s\;
|\epsilon^{UEG-ISI}_c(d,r_s)-\epsilon^{exact}_c(r_s)|,
\end{equation}
and the results are reported in Fig. \ref{f1}. At $d\approx 0$ and $d\approx 0.74$, the
Error(0)$\approx$0.138 and Error(0.74)$\approx$ 0.08
are similar with the ones given by ISI and revISI, respectively. The curve Error$(d)$ has
only one minimum at $d=3.5$ where Error(3.5)$\approx 0.003$.

%
\begin{figure}[hbt]
\includegraphics[width=\columnwidth]{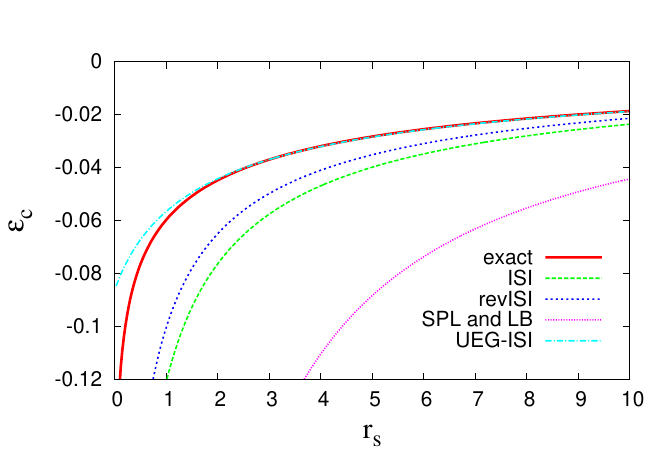}
\caption{ UEG correlation energy per particle $\epsilon_c(r_s)$ versus the bulk
parameter $r_s$, from several ISI-like functionals. The exact curve is the accurate
parametrization of the Perdew and Wang \cite{perdew1992accurate}.
\newline
}
\label{f2}
\end{figure}
%
In  Fig. \ref{f2} we show our UEG results. Both SPL and LB functionals
behave similarly as $\epsilon_c=-0.44196/r_s$, being accurate only in the low-density limit.
On the other hand, the ISI functional is definitely better and in the high-density limit,
behaves as
\begin{equation}
\epsilon_c^{ISI}\longrightarrow -0.1736/\sqrt{r_s}, \;\;\;\;\;\rm{when}\;\;\;\;\; r_s\rightarrow 0.
\label{eq15l3}
\end{equation}
Moreover, the revISI shows a quite good improvement over the ISI functional, behaving as
\begin{equation}
\epsilon_c^{revISI}\longrightarrow -0.130/\sqrt{r_s}, \;\;\;\;\;\rm{when}\;\;\;\;\; r_s\rightarrow
0.
\label{eq16l3}
\end{equation}
However, the UEG-ISI functional shows the best performance, being almost exact for $r_s\ge
1$, while in the high-density limit is just a constant
\begin{equation}
\epsilon_c^{UEG-ISI}\rightarrow -0.0192(1+d)=-0.086, \;\rm{when}\;
r_s\rightarrow 0.
\label{eq16}
\end{equation}
We recall that $\epsilon^{exact}_c\longrightarrow 0.031091 \ln(r_s) - 0.0469203$ for
spin-unpolarized case \cite{perdew1992accurate} when $r_s\rightarrow
0$, such that the exchange energy per particle $\epsilon_x\longrightarrow
-0.4582/r_s$ is dominating. We note also that the constant term in the exact high-density limit
expansion is almost twice smaller that the one from UEG-ISI limit.

\subsection{Jellium clusters}
\label{sec32}

%
\begin{figure}[hbt!]
\includegraphics[width=\columnwidth]{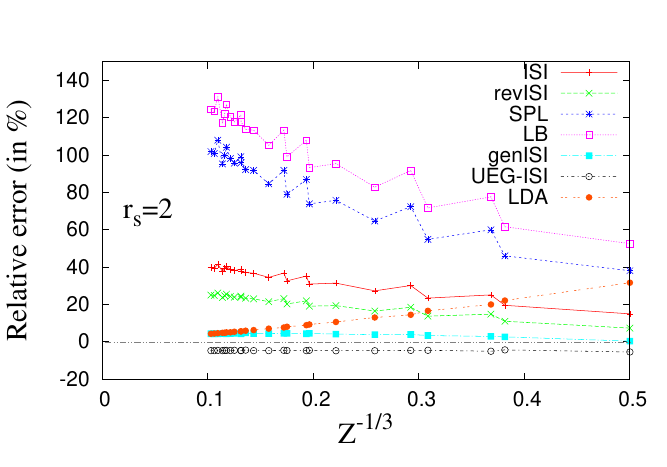}
\includegraphics[width=\columnwidth]{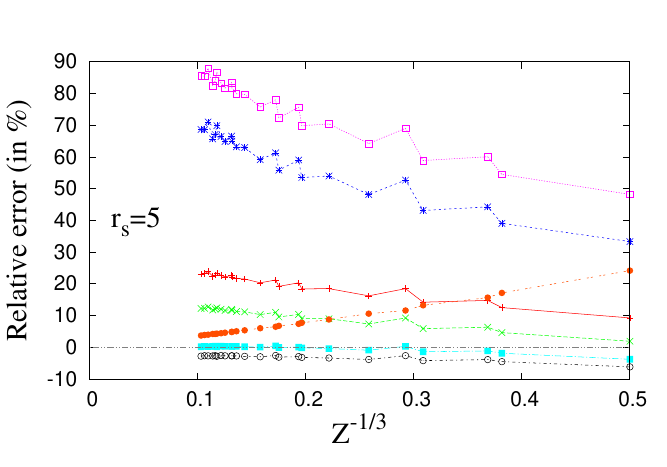}
\caption{Relative error $[(approx-reference)/reference]\times 100$ of the correlation
energy of jellium clusters (from $Z=$8 to 912 $e^-$) versus $Z^{-1/3}$,
for $r_s=2$ (upper panel) and $r_s=5$ (lower panel).
For the reference correlation energy, we use the TPSS one, which is very accurate
for such systems \cite{tao2008nonempirical}.
}
\label{f3}
\end{figure}
%
We present in Fig. \ref{f3} the correlation energy relative errors (RE) of the
considered jellium clusters for $r_s=2$ (upper panel) and $r_s=5$ (lower panel), respectively.
The ISI-like methods have a similar trend as in the UEG case (see Fig. \ref{f1}). revISI gives a good
improvement over the ISI functional, but in both cases, the errors increase with $Z$. Contrary to
this trend, the LDA error decreases with $Z$. However, the best performances are reported from
the genISI (with $0.3\% \le \rm{RE} \le 4.8 \% $ for the upper panel, and $-4\% \le
\rm{RE} \le 0.5 \% $ for the lower panel) and the
UEG-ISI (with $-3.8\% \le \rm{RE} \le -2.8 \% $ for the upper panel, and $-5.6\% \le \rm{RE}
\le -2.2 \% $ for the lower panel).

\begin{table}[htbp]
\caption{Mean absolute relative error (MARE in \%) of correlation energy
from 24 jellium clusters ($8\le Z\le 912$), for several bulk parameters. The best results
have been highlighted in boldface. We use the TPSS
correlation as the reference correlation energy, because it predicts with high accuracy
the diffusion Monte Carlo (DMC) results corrected for the fixed-node DMC error
(see tables VI and VII of Ref. \onlinecite{tao2008nonempirical}).
To our best knowledge, there are not high-level, benchmark results available for
jellium clusters with $Z>106$.
}
\begin{tabular}{lccccccccc}
\hline \hline
 $r_s$       & 2 & 3 & 4 & 5 \\
 \hline\hline
ISI & 33.7 & 27.0 & 22.8 & 19.9 \\
revISI & 20.9 & 15.5 & 12.2 & 10.0 \\
SPL & 83.9 & 72.4& 64.4 & 58.5 \\
LB  & 104.4 & 91.4& 82.1& 75.0 \\
UEG-ISI  & 4.7 &3.8 & 3.4 & 3.1 \\
genISI  & \textbf{4.0} & \textbf{1.8}& \textbf{0.9} & \textbf{0.6} \\
LDA & 9.7 & 8.7 & 8.2 & 7.9 \\
\hline \hline
\end{tabular}
\label{tab1}
\end{table}
We summarize our results in Table \ref{tab1}, where we present the MAREs in correlation energy
of the considered XC functionals, for several values of bulk parameters.
We observe that all functionals show a worsening of their prediction when $r_s$
decreases. However, both genISI and UEG-ISI are remarkably accurate, with MARE$\le 5 \%$.

%
\begin{figure}[hbt!]
\includegraphics[width=\columnwidth]{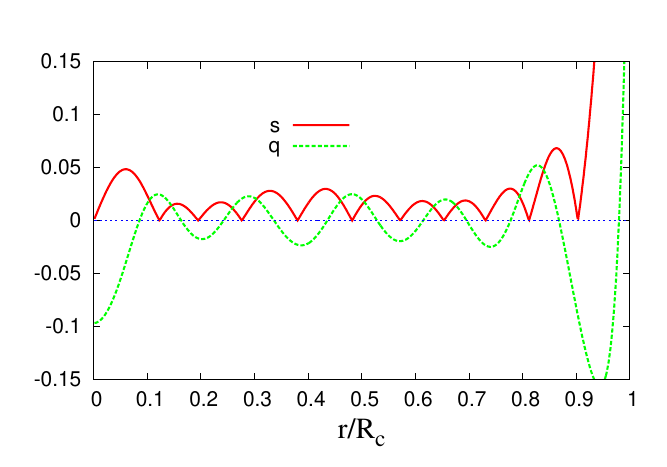}
\includegraphics[width=\columnwidth]{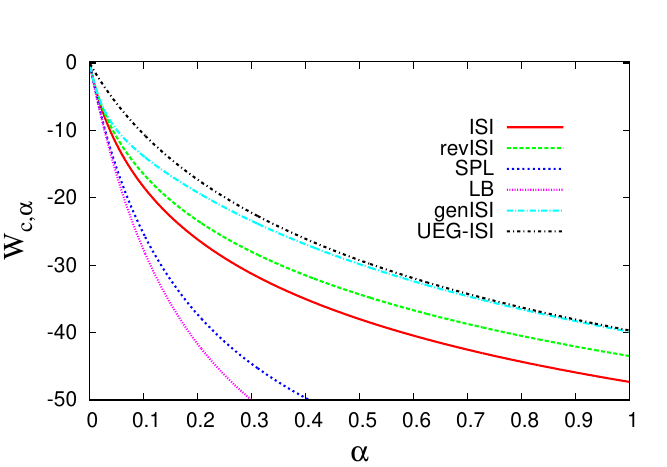}
\caption{Upper panel: The reduced gradient $s=|\nabla n|/[2k_Fn]$ and the reduced Laplacian
$q=\nabla^2 n/[4k^2_Fn]$, versus the
normalized radial distance $r/R_c$, for the jellium sphere with $Z=912$ and $r_s=4$.
Here $R_c=r_s Z^{1/3}\approx 38.79\;$a.u. is the radius of the sphere.
\newline
Lower panel: The adiabatic connection correlation integrand $W_{c,\alpha}=W_{xc,\alpha}-W_0$ versus
$\alpha$ for the same jellium cluster ($Z=912$ and $r_s=4$).
The area under each curve is the
correlation energy: $E_c^{ISI}=-34.856$ Ha, $E_c^{revISI}=-31.605$ Ha, $E_c^{genISI}=-27.759$
Ha,$E_c^{UEG-ISI}=-26.689$ Ha, and $E_c^{TPSS}=-27.525$ Ha.
}
\label{f4}
\end{figure}
%
Finally, let us consider the jellium sphere with $Z=912$ and $r_s=4$. In the upper panel
of Fig. \ref{f4}, we show the reduced density gradients $s(\R)$ and $q(\R)$ inside of this sphere. Note that $q$ in this paragraph is the reduced Laplacian and not the global parameter in Eq. \ref{eq:qdef}. We
observe that both
$s$ and $q$ are very small ($s\le 0.1$, $-0.1\le q\le 0.05$), showing the shell structure
oscillations, and only near the boundary of the cluster, when the density starts to decay,
the reduced gradients become large. Thus, the core of this sphere is a
typical example of a slowly varying density, a difficult case for the ISI-like functionals.
In the lower panel of Fig. \ref{f4}, we report the coupling constant correlation integrand
$W_{c,\alpha}=W_{xc,\alpha}-W_0$, for $0 \le \alpha \le 1$. The area under each curve is the
correlation energy. We observe that genISI has the same correct behavior at $\alpha\rightarrow
0$ as ISI and revISI (see Eq. (\ref{eq4})), while for $\alpha\ge 0.3$, it smoothly recovers
the UEG-ISI functional.
\subsection{Noble atoms}
\label{sec33}
%
\begin{figure}[hbt!]
\includegraphics[width=\columnwidth]{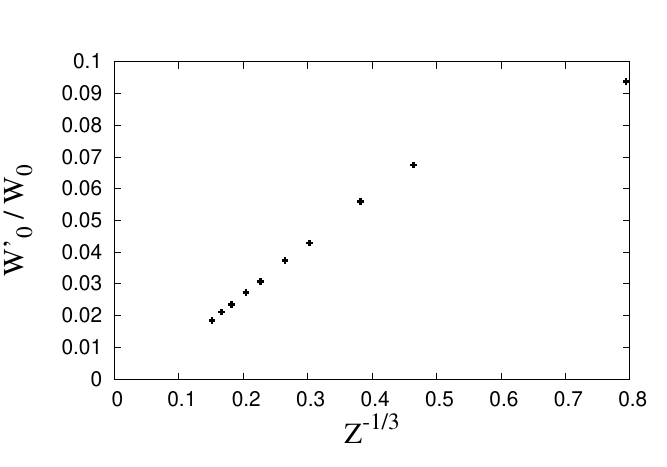}
\includegraphics[width=\columnwidth]{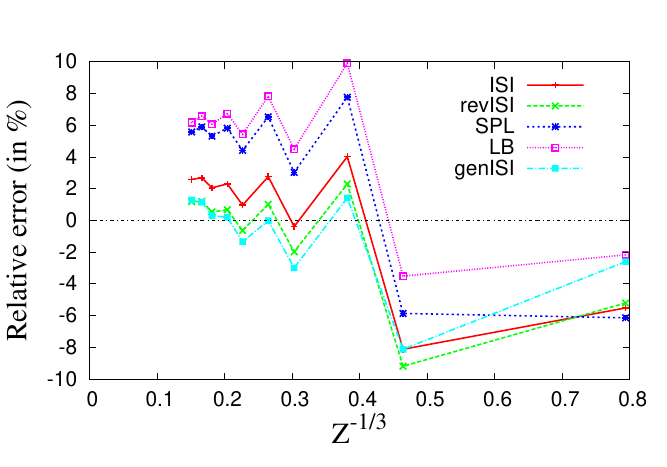}
\caption{Upper panel: The ratio $p[n]=W'_0/W_0$ of noble atoms (from $Z=$2 to 290 $e^-$)
versus $Z^{-1/3}$.
\newline
Lower panel: Relative error $[(approx-reference)/reference]\times 100$ of the correlation
energy of noble atoms (from $Z=$2 to 290 $e^-$) versus $Z^{-1/3}$.
The reference correlation energy is: for $Z\le 86$ (a.i. until Rn atom) we use the benchmark data
shown in Tables I and II of Ref. \onlinecite{burke2016locality} (that was taken from Refs.
\cite{mccarthy2011accurate,mccarthy2012does,chakravorty1993ground}), and for the atoms with $118\le Z
\le 290$ we use the acGGA correlation functional of Ref. \onlinecite{burke2014atomic}, that was built from
semiclassical atom theory, being very accurate for large atoms.
}
\label{f5}
\end{figure}
%
In the upper panel of Fig. \ref{f5}, we report the ratio $p[n]=W'_0/W_0$ of the noble atoms
(from $Z=$2 to 290 $e^-$). One can note that it is very small, decreasing from 0.094 (He atom)
to 0.018 (290 e$^-$ atom),
in an almost linear pattern that predicts $p[n]\rightarrow 0$ in the limit $Z\rightarrow
\infty$. Thus, the
core of large atoms is a typical example of a high-density system where the exchange energy
dominates over the GL2 corelation energy.

In the lower panel of Fig. \ref{f5}, we report the relative errors of the correlation
energy for these systems. All ISI-like functionals are quite accurate, within a maximum error of
$\pm$ 10\%. Note that genISI and revISI give similar results, showing the best performances.
In this case, the UEG-ISI functional, not shown in the figure, fail badly, with relative
errors $-60\% \le \rm{RE} \le 80\%$.

%
\begin{figure}[hbt!]
\includegraphics[width=\columnwidth]{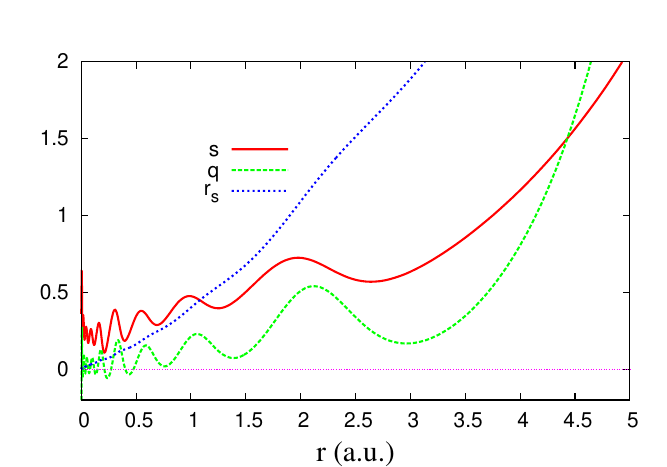}
\includegraphics[width=\columnwidth]{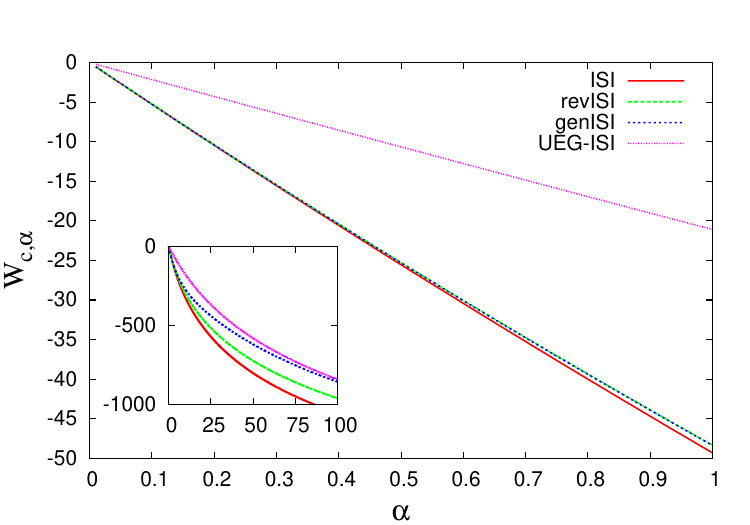}
\caption{Upper panel:
The bulk parameter $r_s(\R)$ and the reduced gradients $s(\R)$ and $q(\R)$,
versus the radial distance from the nucleus $r$, for the noble atom with $Z=290$.
\newline
Lower panel: The adiabatic connection correlation integrand $W_{c,\alpha}$ versus
$\alpha$ ($0\le \alpha \le 1$) for the noble atom with $Z=290$.
The area under each curve is the
correlation energy: $E_c^{ISI}=-25.240$ Ha, $E_c^{revISI}=-24.895$ Ha, $E_c^{genISI}=-24.924$
Ha, $E_c^{UEG-ISI}=-10.613$ Ha, and $E_c^{ref}=-24.602$ Ha.
The inset shows $W_{c,\alpha}$ versus $\alpha$, for $\alpha\le 100$.
}
\label{f6}
\end{figure}
%
In order to understand better the behavior of genISI for large atoms, we investigate in more detail
the case of the noble atom with $Z=290$. In the upper panel of Fig. \ref{f6}, we show that $s$ and
$q$ are small inside the atomic core, with the exception of nucleus region where $q\rightarrow
-\infty$ and $s\approx 0.4$, due to the cusp of the density at the nucleus. However, at the nucleus,
$r_s\approx 0.0024$ is very small, such that this region approaches the high-density limit.

In the lower panel of Fig. \ref{f6}, we report the coupling constant correlation integrand
$W_{c,\alpha}$, for $0 \le \alpha \le 1$, such that the area under each curve is the
correlation energy. We observe that $W^{genISI}_{c,\alpha}$ and $W^{revISI}_{c,\alpha}$ are
almost indiscernible. Now, the UEG-ISI is
very different from genISI, and only at large coupling constants ($\alpha\ge 50$) both
functionals will start to agree (as shown in the inset of the lower panel of Fig. \ref{f6}).
\subsection{Small systems with exact ingredients ($W_0$, $W'_0$, $W^\text{SCE}_\infty$, and  ${W'}^\text{SCE}_\infty$)}
\label{sec34}
\begin{table}[htbp]
\caption{The correlations energies for several ISI-like expressions computed for Hooke’s atom
with force constant $\omega = 0.5$, two-electron exponential density (Exp) with $n(r) = 2\; exp{(-2r)}/\pi$, and He, Be, and Ne atoms using the exact
ingredients ($W_0$, $W'_0$, $W^\mathrm{SCE}_\infty$, and
$W'^{\mathrm{SCE}}_\infty$).
All reference data have been taken from Table I of Ref. \onlinecite{constantin2019correlation} and the references therein. In the last column, we report the MARE (in \%) for a given method.}
\begin{tabular}{ccccccc}
\hline
 & Hook& Exp. & He & Be & Ne  & \\ \hline \hline
$W^\mathrm{SCE}_\infty$ & -0.743 & -0.910 & -1.500 & -4.020 & -20.000 \\
$W'^{\mathrm{SCE}}_\infty$ & 0.208 & 0.308$^{a)}$ & 0.621 & 2.590 & 22.000 \\
$W_0$ & -0.515 & -0.625 & -1.025 & -2.674 & -12.084 \\
$W'_0$ & -0.101 & -0.093 & -0.101 & -0.250 & -0.938 \\
 &  &  &  &  & & MARE (\%) \\ \hline
SPL & -0.036 & -0.035 & -0.042 & -0.106 & -0.420 & 6.0 \\
LB & -0.038 & -0.037 & -0.044 & -0.110 & -0.432 & 6.3 \\
ISI & -0.037 & -0.036 & -0.043 & -0.104 & -0.410 & 4.6 \\
revISI & -0.037 & -0.036 & -0.043 & -0.104 & -0.405 & 4.0 \\
UEG-ISI & -0.061 & -0.064 & -0.086 & -0.144 & -0.474 & 61.1 \\
genISI & -0.040 & -0.038 & -0.043 & -0.108 & -0.411 & 4.5 \\
Exact & -0.039 & -0.037 & -0.042 & -0.096 & -0.394 \\ \hline \hline
\end{tabular} \\
$a)$ - computed using hPC model from Ref.\onlinecite{smiga2022selfconsistent}
\label{tab2}
\end{table}
As a final part of this section, in Table \ref{tab2}, we show the correlation energies for
several ISI-like expressions for two model systems and a few small atoms, using very
accurate low- and high-density expansion ingredients ($W_0$, $W'_0$, $W_\infty$, and
$W'_\infty$). In the last column, we report the MARE for each
method. The inspection of the table reveals that ISI, revISI and genISI have almost the same accuracy with a MARE in the range
4.0\%-4.6\% .
The SPL and LB expressions provide errors, which are about
2\% worse than ISI. The worst performance is given by the UEG-ISI formula,
which yields MARE of 61.1\%. This is not surprising since this model was designed to be
accurate in the UEG limit where  $W'_0 \rightarrow -\infty$, and it shows the importance of the $E_{xc}^{add}$ term.

We also note that the same trend shown in Table \ref{tab2} was also reported for the
correlation
energies of noble atoms with $2\le Z \le 290$ (see the lower panel of Fig. \ref{f5}), even if
they were computed with meta-GGA approximations for $W'_0$, $W_\infty$, and
$W'_\infty$.

\section{Conclusions}
\label{sec4}
We have constructed the UEG-ISI XC functional, which depends on
$W_0[n]$, $W_\infty[n]$ and $W'_\infty[n]$, being remarkably accurate for the UEG model system.
The UEG-ISI XC functional behaves correctly in the strong-interaction limit, while in the
weak-interaction limit it recovers only the leading term $W_0[n]$. We have also developed the
genISI XC functional, by adding a GL2 correction to the UEG-ISI, such that the genISI fulfills
all the known exact requirements, and additionally it is accurate for the UEG, especially in the
region $1\le r_s\le 10$, that is the most relevant in solid-state applications.

The genISI and UEG-ISI have been tested for the correlation energies of 24 neutral jellium
spheres with number of electrons $8\le Z\le 912$.
The $W_{xc,\alpha}^{genISI}$ and
$W_{xc,\alpha}^{UEG-ISI}$ are similar, with the exception of $\alpha \rightarrow 0$ region, where
$W_{xc,\alpha}^{genISI}$
recovers the first two terms of the GL perturbation expansion of Eq. (\ref{eq4}).
Consequently, both genISI and UEG-ISI give very good and almost similar
performances for jellium clusters, as reported in Table \ref{tab1} and Fig. \ref{f3}.

We have also tested the genISI and UEG-ISI functionals for noble atoms with $2\le Z\le 290$,
where the ratio $p[n]$ is small ($p[n]\le 0.1$), such that genISI and UEG-ISI behave
differently. Thus while the genISI is quite accurate with a relative error below 7\% (see Fig.
\ref{f5} (lower panel)), the
UEG-ISI fails badly, with errors larger than 40\%, as shown in Fig. \ref{f6} (lower panel). This
fact shows that the genISI correction to the UEG-ISI plays a vital role in this
functional construction.

Finally, we have also presented the results from few small systems where the ingredients $W_0$, $W'_0$, $W^\mathrm{SCE}_\infty$, and
$W'^{\mathrm{SCE}}_\infty$ are known with high accuracy. The genISI shows a reasonable performance, in line with the other ISI-like methods.

 Application of genISI to molecular systems is straightforward, with the methods and limitations previously discussed in Refs. \onlinecite{fabiano2016interaction,vuckovic2018restoring,smiga2020modified,smiga2022selfconsistent,turbo2023}.
We expect that the method of genISI development can be further used for the construction of
more accurate ISI-like functionals, towards their use in solid-state and material science
calculations.

\section*{Acknowledgments}
L.A.C. and F.D.S. acknowledge the financial support from
ICSC - Centro Nazionale di Ricerca in High Performance
Computing, Big Data and Quantum Computing, funded by European Union - NextGenerationEU - PNRR.
F.D.S acknowledges the financial support from
PRIN project no. 2022LM2K5X, Adiabatic Connection for Correlation in Crystals ($AC^3$).
S.\'S. thanks to the Polish National Science Center for the partial financial support under Grant No. 2020/37/B/ST4/02713.

\begin{appendix}
\section{ISI-like methods}

\subsection{Interaction Strength Interpolation (ISI) functional~\cite{seidl2000simulation}}

\begin{equation}
 W_\alpha^{ISI}=W_\infty+\frac{X}{\sqrt{1+\alpha Y}+Z}
\end{equation}

\begin{equation}
 E_{xc}^{ISI}=W_\infty+\frac{2X}{Y}\Big[\sqrt{1+Y}-1-Z\ln\Big(\frac{\sqrt{1+Y}+Z}{1+Z}\Big)\Big]
\end{equation}
with
\begin{eqnarray}
 &X=\frac{xy^2}{z^2}, Y=\frac{x^2y^2}{z^4}, Z=\frac{xy^2}{z^3}-1&\\
 &x=-4E_c^{GL2},  y=W_\infty', z=E_x-W_\infty&
\end{eqnarray}

\begin{eqnarray}
 F^{ISI}(\myq)&=&2- 2\frac{\ln(1+\myq)}{\myq}  \\
f^{ISI}(x)&=&\frac{1}{1+x}
\end{eqnarray}

\subsection{Revised ISI (revISI) functional~\cite{gori2009electronic}}

\begin{equation}
 W_\alpha^{revISI}=W_\infty+\frac{b(2+c\alpha+2d\sqrt{1+c\alpha})}
{2\sqrt{1+c\alpha}(d+\sqrt{1+c\alpha})^2}
\end{equation}

\begin{equation}
 E_{xc}^{revISI}=W_\infty+\frac{b}{\sqrt{1+c}+d}
\end{equation}
with
\begin{eqnarray}
 b&=&-\frac{8E_c^{GL2}(W_\infty')^2}{(E_x-W_\infty)^2}\\
 c&=&\frac{16(E_c^{GL2}W_\infty')^2}{(E_x-W_\infty)^4}\\
 d&=&-1-\frac{8E_c^{GL2}(W_\infty')^2}{(E_x-W_\infty)^3}
\end{eqnarray}

Now, putting values of $b$,$c$, and $d$ in Eq.(B7) we get
\begin{eqnarray}
    & &E_{xc}^{revISI} = W_{\infty}  \nonumber \\
    &+&\frac{8E_c^{GL2}(W_\infty')^2}{(W_0-W_\infty)^2 (1+\frac{8 E_c^{GL2}(W_\infty')^2}{(W_0-W_\infty)^2}-\sqrt{1+\frac{16(E_c^{GL2}W_\infty')^2}{(W_0-W_\infty)^4}}}\nonumber\\
\end{eqnarray}




\begin{eqnarray}
    F^{revISI}(\myq)&=& \frac{2\myq}{\myq+2} \\
     f^{revISI}(x)&=&  \frac{4+x}{x^2+4 x+4}
\end{eqnarray}

%

\subsection{Seidl-Perdew-Levy (SPL) functional~\cite{seidl1999strictly}}

\begin{equation}
 W_\alpha^{SPL}=W_\infty+\frac{W_0-W_\infty}{\sqrt{1+2\alpha\chi}}
\end{equation}

\begin{equation}
 E_{xc}^{SPL}=E_x+(E_x-W_\infty)\Big[\frac{\sqrt{1+2\chi}-1-\chi}{\chi}\Big]
\end{equation}
with
\begin{equation}
 \chi=\frac{2E_c^{GL2}}{W_\infty-E_x}
\end{equation}

\begin{eqnarray}
F^{SPL}(\myq)=f^{SPL}(x)=0
\end{eqnarray}
\subsection{Liu-Burke (LB) functional~\cite{liu2009adiabatic}}

\begin{equation}
 W_\alpha^{LB}=W_\infty+b(y+y^4)
\end{equation}

\begin{equation}
 E_{xc}^{LB}=E_x+\frac{2b}{c}\Big[\sqrt{1+c}-\frac{1+c/2}{1+c}-c\Big]
\end{equation}
with
\begin{equation}
y=\frac{1}{\sqrt{1+c\alpha}},
 b=\frac{E_x-W_\infty}{2}, c=\frac{8E_c^{GL2}}{5(W_\infty-E_x)}
\end{equation}
\begin{equation}
F^{LB}(\myq)=f^{LB}(x)=0
\end{equation}



\end{appendix}

\twocolumngrid
\bibliography{gisi.bib}
\bibliographystyle{apsrev4-1}

\end{document}